\documentclass[aps,prd,preprint,groupedaddress]{revtex4-1}

\usepackage{amsmath}
\usepackage{graphicx}
\usepackage{amsbsy}
\usepackage{amssymb}
\usepackage{times,avant,graphicx,color}
\usepackage{dcolumn}% Align table columns on decimal point
\usepackage{bm}% bold math
\usepackage{color}
\begin{document}

\title{Viscous-elastic dynamics of power-law fluids within an elastic cylinder}

\author{Evgeniy Boyko}
\author{Moran Bercovici}
\email[Corresponding author: ]{mberco@technion.ac.il}
\author{Amir D. Gat}
\email[Corresponding author: ]{amirgat@technion.ac.il}
\affiliation{Faculty of Mechanical Engineering, Technion - Israel Institute of Technology, Haifa, Israel}
%\affiliation{*Corres author\\Faculty of Mechanical Engineering, Technion - Israel Institute of Technology, Haifa, Israel}

\begin{abstract}
In a wide range of applications, microfluidic channels are implemented in soft substrates. In such configurations, where fluidic inertia and compressibility are negligible, the propagation of fluids in channels is governed by a balance between fluid viscosity and elasticity of the surrounding solid. The viscous-elastic interactions between elastic substrates and non-Newtonian fluids are particularly of interest due to the dependence of viscosity on the state of the system. In this work, we study the fluid-structure interaction dynamics between an incompressible non-Newtonian fluid and a slender linearly elastic cylinder under the creeping flow regime. Considering power-law fluids and applying the thin shell approximation for the elastic cylinder, we obtain a non-homogeneous p-Laplacian equation governing the viscous-elastic dynamics. We present exact solutions for the pressure and deformation fields for various initial and boundary conditions for both shear-thinning and shear-thickening fluids. We show that in contrast to Stokes' problem where a compactly supported front is obtained for shear-thickening fluids, here the role of viscosity is inversed and such fronts are obtained for shear-thinning fluids. Furthermore, we demonstrate that for the case of a step in inlet pressure, the propagation rate of the front has a $t^{\frac{n}{n+1}}$ dependence on time ($t$), suggesting the ability to indirectly measure the power-law index ($n$) of shear-thinning liquids through measurements of elastic deformation. 
\end{abstract}
\maketitle
\section{Introduction}
\vspace{-0.1in}
Viscous-elastic interaction (VEI) is a sub-field of fluid-structure interaction focused on fluid motion at low-Reynolds numbers \cite{DupratStone}. VEI received considerable attention over recent years due to its relevance to a wide spectrum of natural processes and applications. These include biological flows \cite{Ku, Nichols}, geophysical and geological flows \cite{Hewitt2015}, suppression of viscous fingering instabilities \cite{Pihler-Puzovic2012,Pihler-Puzovic2013,AlHousseiny,Pihler-Puzovic2014,Pihler-Puzovic2015,Peng2015}, manufacturing of micro-electro-mechanical systems \cite{Hosoi2004,Lister2013}, development of flexible microfluidic valves \cite{Holmes}, and soft robotics \cite{Shepherd,Martinez,Marchese,Matia}.
\vspace{-0.1in}

In particular, the case of Newtonian viscous flow through elastic channels has been extensively studied as a model problem for internal VEI flows such as those found in veins and arteries \cite{Canic}, collapsible tubes (see \cite{Grotberg,Heil} and \cite{DupratStone}, ch. 8,  and references therein), and soft actuators \cite{Elbaz2014}. In the field of micro and nano-fluidics, channels are often fabricated from flexible materials (\textit{e.g.} PDMS), which deform under internal or external pressures \cite{Hardy,deRutte}. While many microfluidic applications involve non-Newtonian fluids, very few studies examined the deformation field and flow of non-Newtonian fluids within elastic configurations. In addition to an academic interest in such systems, we believe they hold potential for novel applications, particularly in soft robotics and soft lab-on-a-chip systems, where the choice of fluid properties could allow new modes of actuation and flow control. 
\vspace{-0.2in}

In this work, we extend the recent study of Elbaz and Gat \cite{Elbaz2014} which focused on Newtonian fluids, and analyze the viscous-elastic dynamics of non-Newtonian power-law fluids flowing through a slender linearly elastic cylinder. In Sec. II, we present the problem formulation as well as the key assumptions we use in the derivation of the model. In Sec. III we apply the lubrication approximation to the flow equations and the thin shell approximation to the elasticity equations and derive a non-linear p-Laplacian governing equation for the viscous-elastic dynamics. We provide appropriate scaling for physical variables and clarify the range of validity of the model. In Sec. IV we examine the viscous-elastic dynamics of the system, subjected to various initial and boundary conditions, and obtain both exact and numerical solutions for the pressure and deformation fields for both shear-thinning and shear-thickening fluids. In Sec. V we demonstrate and discuss the inverse role of viscosity in the p-Laplacian equation governing the viscous-elastic problem, as compared to Stokes' problems \cite{Pascal1992,Ai,Pritchard}. We conclude with a discussion of the results in Sec. VI. 
\vspace{-0.2in}

\section{Problem formulation}
We study the low-Reynolds number fluid-structure interaction dynamics of non-Newtonian,
incompressible, axisymmetric flow through a slender linearly elastic
cylinder. Figure 1 presents a schematic illustration of the configuration
and coordinate system. We hereafter denote dimensional variables by
tildes, normalized variables without tildes and characteristic values
by an asterisk superscript. 
\begin{figure}[h]
 \centerline{\includegraphics[scale=0.8]{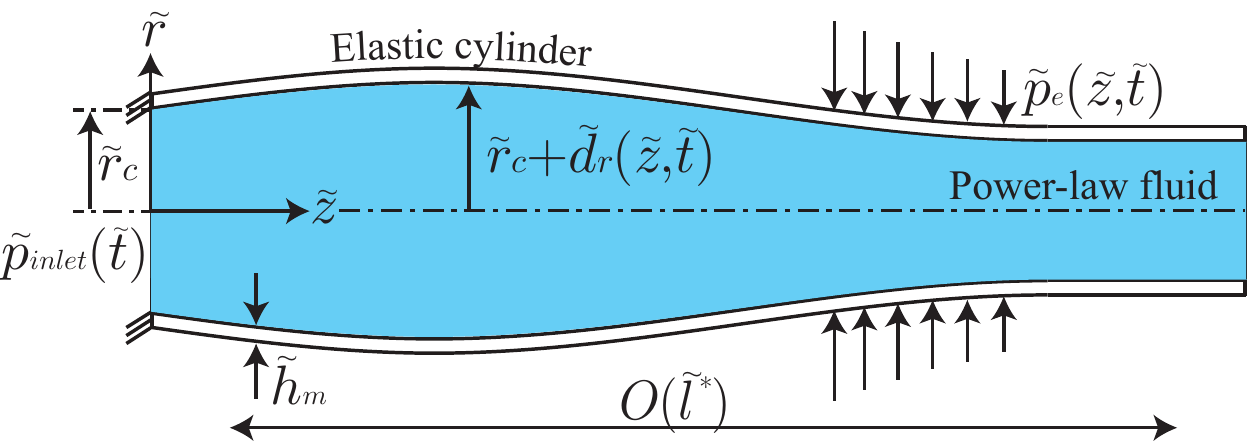}}
\caption{Schematic illustration of the examined configuration, showing the
coordinate system and relevant physical parameters. An elastic cylinder
of inner radius $\tilde{r}_{c}$ and wall thickness $\tilde{h}_{m}$ is
clamped at $\tilde{z}=0$ and contains a non-Newtonian power-law fluid.
Viscous-elastic interaction between the fluid and the elastic structure,
induced by externally applied pressures $\tilde{p}_{inlet}$ and $\tilde{p}_{e}$
at the inlet and the outer boundary respectively, leads to radial, $\tilde{d}_{r}$,
and axial, $\tilde{d}_{z}$, displacement fields of the elastic cylinder.}
\end{figure}
\vspace{-0.15in}

We employ a cylindrical coordinate system $(\tilde{r},\tilde{\theta},\tilde{z})$
whose $\tilde{z}$ axes coincides with the symmetry line of the cylinder.
The elastic cylinder is clamped at $\tilde{z}=0$, and has an initial uniform radius $\tilde{r}_{c}$, constant wall
thickness $\tilde{h}_{m}$, Young's modulus $\tilde{E}$ and Poisson's
ratio $\nu$. The fluid's density is $\tilde{\rho}$ and is assumed to be constant, its
velocity is $\mathbf{\tilde{\mathrm{\mathit{\boldsymbol{u}}}}}=\left(\tilde{u}_{r},\tilde{u}_{z}\right)$
and its pressure is $\tilde{p}$. The externally applied pressures
$\tilde{p}_{inlet}$ and $\tilde{p}_{e}$ at the inlet and the
outer shell boundary, result in viscous-elastic interaction between
the non-Newtonian fluid and the elastic cylinder, which in turn leads
to both radial, $\tilde{d}_{r}$, and axial, $\tilde{d}_{z}$, deformation
fields. The characteristic length scale in the $\boldsymbol{\hat{z}}$ direction
is denoted by $\tilde{l}^{*}$, while $\tilde{d}_{r}^{*}$ and $\tilde{d}_{z}^{*}$ are the characteristic
radial and axial deformations, respectively.

We further assume a slender geometry, thin walls, negligible fluidic inertia effects (small Womersley number), and small deformations in both the radial and axial directions. These can be summarized as 
\begin{equation}
\epsilon_{1}=\frac{\tilde{r}_{c}}{\tilde{l}^{*}}\ll1,\quad\epsilon_{2}=\frac{\tilde{h}_{m}}{\tilde{r}_{c}}\ll1,\quad Wo\ll1,\quad\epsilon_{3}=\frac{\tilde{d}_{r}^{*}}{\tilde{r}_{c}}\sim\frac{\tilde{d}_{z}^{*}}{\tilde{l}^{*}}\ll1,\label{Epsilon1}
\end{equation}
allowing the use of lubrication and thin elastic shell approximations. The exact expression for the Womersley number appearing in (\ref{Epsilon1}) depends on the viscosity model, which we will define in  (\ref{Reduced Reynolds}). 
Furthermore, defining both ratios, $\tilde{d}_{r}^{*}/\tilde{r}_{c}$ and $\tilde{d}_{z}^{*}/\tilde{l}^{*}$, to be of scale $\epsilon_{3}$ emanates from mass conservation of the fluid and the constitutive law of the elastic structure, as shown in the following section.

\section{Viscous-elastic governing equations for power-law fluids}

\subsection{Fluidic problem}

Description of the non-Newtonian fluid rheology requires the use of
a specific constitutive model. Of all models, the commonly used power-law
model approximately describes the behavior of shear-dependent non-Newtonian
fluids over an intermediate range of shear rates, and at the same
time is sufficiently simple to allow analytical treatment. Therefore,
throughout this work we consider the following constitutive model
for the viscosity $\tilde{\eta}$ \cite{Bird}
\begin{equation}
\tilde{\eta}(\dot{\gamma})=\tilde{\mu}_{eff}\dot{\gamma}^{n-1},\label{Power-law fluid}
\end{equation}
where $n$ is the dimensionless power-law index, $\tilde{\mu}_{eff}$
is the effective constant viscosity with units $\mathrm{Pa\,s^{n}}$
and $\dot{\gamma}$ is the shear rate defined as $\dot{\gamma}=\sqrt{2\boldsymbol{\tilde{D}}:\boldsymbol{\tilde{D}}},$
where $\boldsymbol{\tilde{D}}$ is the rate of deformation tensor
given by $\boldsymbol{\tilde{D}}=(\boldsymbol{\tilde{\nabla}\tilde{u}}+\boldsymbol{\tilde{\nabla}\tilde{u}^{\dagger}})/2.$
The power-law index lies in the range $0<n<1$ for shear-thinning fluids,
and $n>1$ for shear-thickening fluids, while $n=1$ represents the
case of a Newtonian fluid.

Based on the assumptions mentioned above, the relevant governing equations
and boundary conditions are the continuity equation
\begin{equation}
\frac{1}{\tilde{r}}\frac{\partial}{\partial\tilde{r}}\left(\tilde{r}\tilde{u}_{r}\right)+\frac{\partial\tilde{u}_{z}}{\partial\tilde{z}}=0,\label{Continuity}
\end{equation}
the momentum equation in the $\boldsymbol{\hat{z}}$ direction
\begin{equation}
\tilde{\rho}\left(\frac{\partial\tilde{u}_{z}}{\partial\tilde{t}}+\tilde{u}_{r}\frac{\partial\tilde{u}_{z}}{\partial\tilde{r}}+\tilde{u}_{z}\frac{\partial\tilde{u}_{z}}{\partial\tilde{z}}\right)=-\frac{\partial\tilde{p}}{\partial\tilde{z}}+\frac{\partial}{\partial\tilde{z}}\left(\tilde{\tau}_{zz}\right)+\frac{1}{\tilde{r}}\frac{\partial}{\partial\tilde{r}}\left(\tilde{r}\tilde{\tau}_{rz}\right),\label{z momentum}
\end{equation}
the momentum equation in the $\boldsymbol{\hat{r}}$ direction
\begin{equation}
\tilde{\rho}\left(\frac{\partial\tilde{u}_{r}}{\partial t}+\tilde{u}_{r}\frac{\partial\tilde{u}_{r}}{\partial\tilde{r}}+\tilde{u}_{z}\frac{\partial\tilde{u}_{r}}{\partial\tilde{z}}\right)=-\frac{\partial\tilde{p}}{\partial\tilde{r}}+\frac{1}{\tilde{r}}\frac{\partial}{\partial\tilde{r}}\left(\tilde{r}\tilde{\tau}_{rr}\right)+\frac{\partial}{\partial\tilde{z}}\left(\tilde{\tau}_{zr}\right)-\frac{\tilde{\tau}_{\theta\theta}}{\tilde{r}},\label{r momemtum}
\end{equation}
the constitutive equation for the stress tensor $\boldsymbol{\tilde{\tau}}$ of power-law fluid
\begin{equation}
\tilde{\boldsymbol{\tau}}=2\tilde{\mu}_{eff}\dot{\gamma}^{n-1}\tilde{\boldsymbol{D}},\label{Stress tensor}
\end{equation}
as well as the no-slip and the no-penetration conditions at the solid-liquid interface 
\begin{equation}
\tilde{u}_{z}|_{\tilde{r}=\tilde{r}_{c}+\tilde{d}_{r}}=\frac{\partial\tilde{d_{z}}}{\partial\tilde{t}},\quad\tilde{u}_{r}|_{\tilde{r}=\tilde{r}_{c}+\tilde{d}_{r}}=\frac{\partial\tilde{d_{r}}}{\partial\tilde{t}}+\tilde{u}_{z}|_{\tilde{r}=\tilde{r}_{c}+\tilde{d}_{r}}\frac{\partial d_{r}}{\partial z},\label{Boundary conditions}
\end{equation}
where $\tilde{t}$ is the time.

Scaling by the characteristic dimensions, we define the normalized
coordinates, $(r,z)=(\tilde{r}/\tilde{r}_{c},\tilde{z}/\tilde{l}^{*})$,
normalized velocity, $(u_{r},u_{z})=(\tilde{u}_{r}/\tilde{u}_{r}^{*},\tilde{u}_{z}/\tilde{u}_{z}^{*})$,
normalized pressure, $p=\tilde{p}/\tilde{p}^{*}$, and normalized
time, $t=\tilde{t}/\tilde{t}^{*}$, where $\tilde{t}^{*}$ is a characteristic
time scale, $\tilde{p}^{*}$ is a characteristic pressure, $\tilde{u}_{r}^{*}$
and $\tilde{u}_{r}^{*}$ are radial and axial characteristic velocities,
respectively. 

Substituting the normalized parameters into the continuity equation (\ref{Continuity}), order
of magnitude analysis yields the balance between radial and axial velocities 
\begin{equation}
\frac{\tilde{u}_{r}^{*}}{\tilde{u}_{z}^{*}}\sim\frac{\tilde{r}_{c}}{\tilde{l}^{*}}=\epsilon_{1},\label{Continuity-scaling}
\end{equation}
and suggests the scaling for shear rate, $\dot{\gamma}$ ,
\begin{equation}
\dot{\gamma}\sim\frac{\tilde{u}_{z}^{*}}{\tilde{r}_{c}}\left(\left|\frac{\partial u_{z}}{\partial r}\right|+O(\epsilon_{1}^{2})\right).\label{Approximation  for shear rate}
\end{equation}
Order of magnitude analysis of the boundary conditions (\ref{Boundary conditions}) yields
the characteristic viscous-elastic times scale
\begin{equation}
\tilde{t}^{*}\sim\frac{\tilde{d}_{z}^{*}}{\tilde{u}_{z}^{*}}\sim\frac{\tilde{d}_{r}^{*}}{\tilde{u}_{r}^{*}}.\label{Time scale}
\end{equation}
Substituting (\ref{Continuity-scaling}) into (\ref{z momentum})
and performing order of magnitude analysis, we obtain the axial characteristic
velocity in terms of the characteristic pressure 
\begin{equation}
\tilde{u}_{z}^{*}=\frac{\epsilon_{1}^{1+1/n}\tilde{p}^{*1/n}\tilde{l}^{*}}{\tilde{\mu}_{eff}^{1/n}},\label{Typical velocity}
\end{equation}
as well as the condition for a negligible Womersley number in terms
of relevant physical quantities
\begin{equation}
Wo=\frac{\tilde{\rho}\tilde{r}_{c}^{2}}{\tilde{\eta}(\dot{\gamma})\tilde{t}^{*}}=\frac{\epsilon_{1}}{\epsilon_{3}}\frac{\tilde{\rho}\tilde{u}_{z}^{*2-n}\tilde{r}_{c}^{n}}{\tilde{\mu}_{eff}}=\frac{\tilde{\rho}\tilde{u}_{z}^{*2-n}\tilde{r}_{c}^{n+1}}{\tilde{\mu}_{eff}d_{z}^{*}}\ll1,\label{Reduced Reynolds}
\end{equation}
representing the ratio between the inertial-viscous time scale $\tilde{\rho}\tilde{r}_{c}^{2}/\tilde{\eta}(\dot{\gamma})=\tilde{\rho}\tilde{r}_{c}^{2}/\tilde{\mu}_{eff}(\tilde{u}_{z}^{*}/\tilde{r}_{c})^{n-1}$
and the viscous-elastic time scale $\tilde{t}^{*}=\tilde{d}_{z}^{*}/\tilde{u}_{z}^{*}.$

Substituting (\ref{Continuity-scaling})-(\ref{Reduced Reynolds})
into (\ref{Continuity})-(\ref{Boundary conditions}) and applying
the lubrication approximation results in the following set of normalized
equations and boundary conditions, 
\begin{equation}
\frac{1}{r}\frac{\partial}{\partial r}\left(ru_{r}\right)+\frac{\partial u_{z}}{\partial z}=0,\label{Continuity Final}
\end{equation}
\begin{equation}
\frac{\partial p}{\partial z}=\frac{1}{r}\frac{\partial}{\partial r}\left(r\left|\frac{\partial u_{z}}{\partial r}\right|^{n-1}\frac{\partial u_{z}}{\partial r}\right)+O(Wo,\epsilon_{1}^{2}),\label{Momentum z-Final}
\end{equation}
\begin{equation}
\frac{\partial p}{\partial r}=O(\epsilon_{1}^{2}Wo,\epsilon_{1}^{2}),\label{Momentum r-Final}
\end{equation}
and
\begin{equation}
u_{z}|_{r=1+\epsilon_{3}d_{r}}=\frac{\partial d{}_{z}}{\partial t},\quad u_{r}|_{r=1+\epsilon_{3}d_{r}}=\frac{\partial d_{r}}{\partial t}+\epsilon_{3}\frac{\partial d{}_{z}}{\partial t}\frac{\partial d_{r}}{\partial z}.\label{Boundary conditions-final}
\end{equation}
Integrating (\ref{Momentum z-Final}) with respect to $r$ and applying
the symmetry condition at $r=0$ yields
\begin{equation}
\frac{\partial u_{z}}{\partial r}=\left|\frac{\partial p}{\partial z}\right|^{\frac{1}{n}-1}\frac{\partial p}{\partial z}\left(\frac{r}{2}\right)^{\frac{1}{n}}.\label{du}
\end{equation}
Integrating (\ref{du}) again with respect to $r$ and applying the
no-slip boundary conditions (\ref{Boundary conditions-final}), results in 
\begin{equation}
u_{z}=\frac{1}{2^{\frac{1}{n}}}\frac{n}{n+1}\left|\frac{\partial p}{\partial z}\right|^{\frac{1}{n}-1}\frac{\partial p}{\partial z}\left(r^{1+\frac{1}{n}}-(1+\epsilon_{3}d_{r}(z,t))^{1+\frac{1}{n}}\right)+\frac{\partial d_{z}}{\partial t}.\label{u in z direction}
\end{equation}
Defining volume flux as $q=2\pi\int_{r=0}^{r=1+\epsilon_{3}d_{r}}ru_{z}dr$,
from (\ref{u in z direction}) it follows
\begin{equation}
\frac{q}{2\pi}=\int_{r=0}^{r=1+\epsilon_{3}d_{r}}ru_{z}dr=-\frac{1}{2^{\frac{1+n}{n}}}\frac{n}{3n+1}\left|\frac{\partial p}{\partial z}\right|^{\frac{1}{n}-1}\frac{\partial p}{\partial z}+\frac{1}{2}\frac{\partial d_{z}}{\partial t}+O(\epsilon_{3}).\label{flux}
\end{equation}
Utilizing (\ref{Boundary conditions-final}) and (\ref{flux}), and
then integrating (\ref{Continuity Final}) with respect to $r$,
we relate the pressure and deformation
fields
\begin{equation}
\frac{\partial d_{r}}{\partial t}+\frac{\partial}{\partial z}\left(-\frac{1}{2^{\frac{1+n}{n}}}\frac{n}{3n+1}\left|\frac{\partial p}{\partial z}\right|^{\frac{1}{n}-1}\frac{\partial p}{\partial z}+\frac{1}{2}\frac{\partial d_{z}}{\partial t}\right)=0.\label{Final governing fluid equation}
\end{equation}

\subsection{Elastic problem}

A similar elastic configuration was considered by Elbaz and Gat \cite{Elbaz2014}, who provided relations between the pressure and deformation fields. For completeness, we here present the governing equations and these key relations, adapted to our notation.  

Focusing on viscous-elastic time scales which are typically significantly longer than elastic inertial time scales, and further neglecting body forces, the deformation field of an axisymmetric linearly elastic material is described by the momentum equations
in the $\boldsymbol{\hat{r}}$ and $\boldsymbol{\hat{z}}$ directions \cite[see][]{Timoshenko,Howell},
\begin{equation}
\frac{\partial}{\partial\tilde{r}}\left(\tilde{r}\tilde{\sigma}_{rr}\right)+\frac{\partial}{\partial\tilde{z}}\left(\tilde{r}\tilde{\sigma}_{rz}\right)-\tilde{\sigma}_{\theta\theta}=0,\quad\frac{\partial}{\partial\tilde{r}}\left(\tilde{r}\tilde{\sigma}_{rz}\right)+\frac{\partial}{\partial\tilde{z}}\left(\tilde{r}\tilde{\sigma}_{zz}\right)=0,\label{r and z momemtum-Elasticity}
\end{equation}
the strain-displacement relations for small deformations,
\begin{equation}
\tilde{e}_{rr}=\frac{\partial\tilde{d}_{r}}{\partial\tilde{r}},\quad\tilde{e}_{\theta\theta}=\frac{\tilde{d}_{r}}{\tilde{r}},\quad\tilde{e}_{zz}=\frac{\partial\tilde{d}_{z}}{\partial\tilde{z}},\quad\tilde{e}_{rz}=\frac{1}{2}\left(\frac{\partial\tilde{d}_{r}}{\partial\tilde{z}}+\frac{\partial\tilde{d}_{z}}{\partial\tilde{r}}\right),\label{Strain}
\end{equation}
and Hooke's law,\begin{subequations}
\begin{equation}
\tilde{e}_{rr}=\frac{1}{\tilde{E}}\left(\tilde{\sigma}_{rr}-\nu\left(\tilde{\sigma}_{zz}+\tilde{\sigma}_{\theta\theta}\right)\right),\quad\tilde{e}_{\theta\theta}=\frac{1}{\tilde{E}}\left(\tilde{\sigma}_{\theta\theta}-\nu\left(\tilde{\sigma}_{zz}+\tilde{\sigma}_{rr}\right)\right),\label{Hooke rr and theta-theta}
\end{equation}
\begin{equation}
\tilde{e}_{zz}=\frac{1}{\tilde{E}}\left(\tilde{\sigma}_{zz}-\nu\left(\tilde{\sigma}_{\theta\theta}+\tilde{\sigma}_{rr}\right)\right),\quad\tilde{e}_{rz}=\frac{1+\nu}{\tilde{E}}\tilde{\sigma}_{rz}.\label{Hooke zz and rz}
\end{equation}
\end{subequations}
The governing equations (\ref{r and z momemtum-Elasticity}) are subjected to the following boundary conditions applied by the liquid 
\begin{equation}
\tilde{\sigma}_{rr}|_{\tilde{r}=\tilde{r}_{c}+\tilde{d}_{r}}=-\tilde{p}+2\tilde{\mu}_{eff}\dot{\gamma}^{n-1}\frac{\partial\tilde{u}_{r}}{\partial\tilde{r}},\quad\tilde{\sigma}_{rz}|_{\tilde{r}=\tilde{r}_{c}+\tilde{d}_{r}}=\tilde{\mu}_{eff}\dot{\gamma}^{n-1}\left(\frac{\partial\tilde{u}_{z}}{\partial\tilde{r}}+\frac{\partial\tilde{u}_{r}}{\partial\tilde{z}}\right),\label{Boundary condition elasticity-1}
\end{equation}
and the external stress 
\begin{equation}
\tilde{\sigma}_{rr}|_{\tilde{r}=\tilde{r}_{c}+\tilde{d}_{r}+\tilde{h}_{m}}=-\tilde{p}_{e},\quad\tilde{\sigma}_{rz}|_{\tilde{r}=\tilde{r}_{c}+\tilde{d}_{r}+\tilde{h}_{m}}=0.\label{Boundary condition elasticity-2}
\end{equation}
We follow the derivation of Elbaz and Gat \cite{Elbaz2014} and obtain the normalized governing equations for the radial and axial deformations in terms of the pressure  
\begin{equation}
d_{r}(z,t)=\left(1-\frac{\nu}{2}\right)p(z,t)-p_{e},\quad d_{z}(z,t)=\int_{0}^{z}\left[(\frac{1}{2}-\nu)p(z,t)+\nu p_{e}\right]dz,\label{Normalized d_r and d_z}
\end{equation}
as well as the characteristic radial deformation, $\tilde{d}_{r}^{*}$,
in terms of characteristic pressure
\begin{equation}
\frac{\tilde{d}_{r}^{*}}{\tilde{r}_{c}}=\frac{\tilde{p}^{*}}{\epsilon_{2}\tilde{E}}\ll1.\label{relation between dr and p}
\end{equation}
In the Supplemental Material \cite{SI}, we show that early scaling of the radial differential, $d\tilde{r}$, in (\ref{r and z momemtum-Elasticity}) by $\tilde{h}_{m}$ (instead of $\tilde{r}_{c}$ \citep[see][]{Elbaz2014}) is a more natural choice which leads to significant simplification in the derivation of (\ref{Normalized d_r and d_z}) and (\ref{relation between dr and p}) in the current configuration.

\subsection{P-Laplacian equation governing the pressure field and its asymptotic approximation}

To combine the fluidic and elastic problems, we substitute elastic relation (\ref{Normalized d_r and d_z})
into the fluidic relation (\ref{Final governing fluid equation}), resulting in a single equation in terms of the pressure 
\begin{equation}
\frac{\partial p}{\partial t}-\frac{\partial}{\partial z}\left(\left|\frac{\partial p}{\partial z}\right|^{\frac{1}{n}-1}\frac{\partial p}{\partial z}\right)=\frac{4-2\nu}{5-4\nu}\frac{\partial p_{e}}{\partial t},\label{Evolution eqaution for the pressure}
\end{equation}
subjected to the following initial and boundary conditions
\begin{equation}
p(z,t=0)=p_{initial}(z),\quad p(z=0,t)=p_{inlet}(t),\label{General BC for p}
\end{equation}
where, for the convenience, we have rescaled the time, $t$, according to
\begin{equation}
t\rightarrow\frac{t}{\alpha},\quad\alpha\triangleq\frac{1}{5-4\nu}\frac{1}{2^{\frac{1-n}{n}}}\frac{n}{3n+1}.\label{Time rescaling}
\end{equation}
Equation (\ref{Evolution eqaution for the pressure}) is a non-homogeneous
one dimensional p-Laplacian equation describing the pressure
in an elastic cylinder containing a power-law fluid subjected to external forcing. 

Similar p-Laplacian equations are encountered in other problems involving the use of non-Newtonian power-law fluids. For example, the p-Laplacian equation represents balance between compressibility and viscosity for the flow of a weakly compressible non-Newtonian fluid in a porous medium \cite{Pascal1985,Pascal1991,DiFederico}, as well as a balance between viscosity and inertia in Stokes' first \cite{Pascal1992,Duffy} and second problems \cite{Ai,Pritchard}, where the operator is applied to the fluid velocity. We here rely on the previous mathematical analysis of the p-Laplacian equation to provide insight into our viscous-elastic problem, such as self-similarity and compact support of the solutions for pressure and deformation fields.

For the case of a non-Newtonian power-law fluid that exhibits a weak
shear-thinning or shear-thickening behavior, we follow Ross, Wilson and Duffy \cite{Ross} and define the auxiliary
small parameter $\varepsilon$, 
\begin{equation}
n=1-\varepsilon,\label{Asym-n}
\end{equation}
and the asymptotic expansion
\begin{equation}
p=p^{(0)}+\varepsilon p^{(1)}+O(\varepsilon^{2}),\label{Asym p}
\end{equation}
where $\left|\varepsilon\right|\ll1$ is a small parameter which is
positive for shear-thinning and negative for shear-thickening behaviors. Using (\ref{Asym-n}) and (\ref{Asym p}) the expansion
$1/n\sim1+\varepsilon+O(\varepsilon^{2})$, yields 
\begin{equation}
\left|\frac{\partial p}{\partial z}\right|^{\frac{1}{n}-1}=1+\varepsilon\ln\left|\frac{\partial p^{(0)}}{\partial z}\right|+O(\varepsilon^{2}).\label{Asym expansion of viscosity}
\end{equation}
We note that the expansion (\ref{Asym expansion of viscosity}) is valid only provided the leading-order pressure gradient is non-zero and bounded throughout the domain.
Utilizing (\ref{Asym-n})-(\ref{Asym expansion of viscosity}), we
obtain the leading-order and first-order corrections of the governing equation (\ref{Evolution eqaution for the pressure}) 
\begin{equation}
O(1):\quad\frac{\partial p^{(0)}}{\partial t}-\frac{\partial^{2}p^{(0)}}{\partial z^{2}}=\frac{4-2\nu}{5-4\nu}\frac{\partial p_{e}}{\partial t},\label{Leading order pressure}
\end{equation}
and
\begin{equation}
O(\varepsilon):\quad\frac{\partial p^{(1)}}{\partial t}-\frac{\partial^{2}p^{(1)}}{\partial z^{2}}=\frac{\partial}{\partial z}\left(\ln\left|\frac{\partial p^{(0)}}{\partial z}\right|\frac{\partial p^{(0)}}{\partial z}\right).\label{First order pressure}
\end{equation}
Both (\ref{Leading order pressure}) and (\ref{First order pressure})
are parabolic equations where the inhomogeneous part of (\ref{Leading order pressure})
is related to the external forcing \citep[see][]{Elbaz2014}, and the inhomogeneous
part of (\ref{First order pressure}) emanates from the non-Newtonian
response of the fluid to the leading-order pressure gradients.

\subsection{Characteristic physical parameters and model range of validity}

We here summarize the characteristic values of relevant physical parameters and the range of validity of the current model.  From
order of magnitude analysis, the characteristic length scale $\tilde{l}^{*}$, the characteristic axial
velocity $\tilde{u}_{z}^{*}$, and the characteristic deformations
$\tilde{d}_{r}^{*}$ and $\tilde{d}_{z}^{*}$ are given by\begin{subequations}
\begin{equation}
\tilde{l}^{*}=\left(\alpha\frac{\tilde{E}\tilde{h}_{m}\tilde{t}^{*}}{\tilde{p}^{*}}\right)^{\frac{n}{n+1}}\left(\frac{\tilde{p}^{*}\tilde{r}_{c}}{\tilde{\mu}_{eff}}\right)^{\frac{1}{n+1}},\label{Characteristic length F}
\end{equation}
\begin{equation}
\tilde{u}_{z}^{*}=\tilde{r}_{c}\left(\frac{\tilde{p}^{*}}{\tilde{\mu}_{eff}}\frac{\tilde{r}_{c}}{\tilde{l}^{*}}\right)^{\frac{1}{n}},\label{Characteristic velocity F}
\end{equation}
\begin{equation}
\tilde{d}_{r}^{*}=\tilde{r}_{c}\frac{\tilde{r}_{c}}{\tilde{h}_{m}}\frac{\tilde{p}^{*}}{\tilde{E}},\label{Characteristic radial deformation F}
\end{equation}
and
\begin{equation}
\tilde{d}_{z}^{*}=\tilde{l}^{*}\frac{\tilde{r}_{c}}{\tilde{h}_{m}}\frac{\tilde{p}^{*}}{\tilde{E}}.\label{Characteristic axial deformation F}
\end{equation}
\end{subequations}
 Since in this work we focus on semi-infinite configuration,
there is no inherent axial length scale to the problem and thus scaling
analysis provides only a relation between $\tilde{l}^{*}$ and $\tilde{t}^{*}$.
We define the characteristic pressure $\tilde{p}^{*}$ and the characteristic
time scale $\tilde{t}^{*}$ from the boundary condition $\tilde{p}(0,\tilde{t})=\tilde{p}_{inlet}(\tilde{t})$
with characteristic amplitude $\tilde{p}_{0}$ and time scale $\tilde{t}_{0}$,
by setting $\tilde{p}^{*}=\tilde{p}_{0}$ and $\tilde{t}^{*}=\tilde{t}_{0}$. 

Substituting (\ref{Characteristic length F})-(\ref{Characteristic axial deformation F})
into (\ref{Epsilon1}) provides formulation of the assumptions made in the model in terms of known physical and geometrical parameters: 
\begin{subequations}
\begin{equation}
\epsilon_{1}=\tilde{r}_{c}\left(\alpha\frac{\tilde{E}\tilde{h}_{m}\tilde{t}_{0}}{\tilde{p}_{0}}\right)^{-\frac{n}{n+1}}\left(\frac{\tilde{p}_{0}\tilde{r}_{c}}{\tilde{\mu}_{eff}}\right)^{-\frac{1}{n+1}}\ll1,\label{Epsilon1 2}
\end{equation}
\begin{equation}
\epsilon_{2}=\frac{\tilde{h}_{m}}{\tilde{r}_{c}}\ll1,\label{Epsilon2 2}
\end{equation}
\begin{equation}
\epsilon_{3}=\frac{\tilde{d}_{r}^{*}}{\tilde{r}_{c}}=\frac{\tilde{d}_{z}^{*}}{\tilde{l}^{*}}=\frac{\tilde{r}_{c}}{\tilde{h}_{m}}\frac{\tilde{p}_{0}}{\tilde{E}}\ll1,\label{Epsilon3 2}
\end{equation}
\begin{equation}
Wo=\frac{\tilde{\rho}\tilde{r}_{c}^{2}}{\tilde{\mu}_{eff}\tilde{t}_{0}}\left(\alpha\frac{\tilde{h}_{m}}{\tilde{r}_{c}}\frac{\tilde{E}}{\tilde{p}_{0}}\frac{\tilde{t}_{0}\tilde{\mu}_{eff}}{\tilde{p}_{0}}\right)^{\frac{n-1}{n+1}}\ll1.\label{Wo 2}
\end{equation}
\end{subequations}  
\begin{table}
  \begin{center}
  \begin{tabular}{lcccc}
%\multicolumn{5}{c}{Characteristic values of physical parameters} \\
Property  &   Symbol & Value & Units & Based on \\
\hline
Wall thickness  &   $\tilde{h}_{m}$ &  50 & $\mu m$ & input\\
Inner radius  & $\tilde{r}_{c}$ & 500 & $\mu m$ & input\\
Young's modulus (PDMS)  & $\tilde{E}$ &  $1\times10^{6} $ & Pa & input\\
Poisson's ratio & $\nu$ &  0.3 & --& input\\
Power-law index (Xanthan suspension) & $n$ &  $0.55$ &--& input \\
Effective viscosity (Xanthan suspension) & $\tilde{\mu}_{eff}$ &0.57 & \,Pa\,s$^{n}$& input\\
Characteristic time scale &  $\tilde{t}^{*}$ & 0.5 & s& input\\
Characteristic pressure &  $\tilde{p}^{*}$ &  $1\times10^{4} $ & Pa & input\\
Axial length scale & $\tilde{l}^{*}$ & $14$ &cm & (\ref{Characteristic length F})\\
Typical axial velocity & $\tilde{u}_{z}^{*}$ & 0.91  &   $\mathrm{m\,s^{-1}}$ &(\ref{Characteristic velocity F})\\
Typical radial deformation & $\tilde{d}_{r}^{*}$ &  50 & $\mu m$&(\ref{Characteristic radial deformation F})\\
Typical axial deformation & $\tilde{d}_{z}^{*}$ & 14 & mm & (\ref{Characteristic axial deformation F})\\
\textbf{Slenderness} & $\epsilon_{1}$ & $3.5\times10^{-3} $ &--& (\ref{Epsilon1 2})\\
\textbf{Wall thinness} & $\epsilon_{2}$ & 0.1 &--& (\ref{Epsilon2 2})\\
\textbf{Smallness of deformations} & $\epsilon_{3}$ & 0.1&--& (\ref{Epsilon3 2}) \\
\textbf{Womersley number} & $Wo$ & $2.5\times10^{-2}$ &--& (\ref{Wo 2})\\
      \end{tabular}
  \caption{Characteristic values of physical parameters corresponding to the flow of Xanthan gum suspension in a 500 $\mu m$ radius PDMS channel, subjected to external pressure of $1\times10^{4} $ Pa. Resulting values for the non-dimensional parameters (in bold) indicate that the model holds well in this regime.}
  \label{T1}
  \end{center}
\end{table}
\vspace{-0.2in}

To illustrate the applicability of these assumptions in soft-robotic and lab-on a-chip applications, we examine a typical micro-configuration made of PDMS \citep[polydimethylsiloxane, $\tilde{E}=10^{6}\,\mathrm{Pa}$,][]{Cheng}
and containing a suspension of Xanthan gum behaving as power-law viscous
fluid \cite[$n=0.55,\,\tilde{\mu}_{eff}=0.57\,$$\mathrm{Pa\,s^{n}}$,][]{Longo}.
Table \ref{T1} presents the characteristic values of physical and geometric
parameters, readily satisfying the assumptions used in the current analysis.

\vspace{-0.2in}
\section{Results}
\vspace{-0.2in}
In this section, we examine the viscous-elastic dynamics resulting from time-varying inlet pressure. We first obtain the pressure field by solving the governing equation (\ref{Evolution eqaution for the pressure}) for arbitrary
values of $n$. Then, using (\ref{flux})
and (\ref{Normalized d_r and d_z}), we provide the corresponding volume
flux and deformation field. 

\subsection{Instantaneous injection of mass at the inlet}

We examine the viscous-elastic dynamics due to an initial localized pressure at $z=0$, which may be represented by 
\begin{equation}
p(z,t=0)=\delta(z),\label{IC impulse}
\end{equation}
where $\delta$ is Dirac's delta function. We note that (\ref{IC impulse})
physically results from instantaneous release of a fluid mass, $\tilde{m}$,
into the elastic cylinder at the inlet at $t=0$ after which ($t>0$), the inlet is sealed and thus
\begin{equation}
\frac{\partial p}{\partial z}\left(z=0,t>0\right)=0.\label{Sealed impulse}
\end{equation}
Mass is thus conserved globally according to 
\begin{equation}
\int_{0}^{\infty}d_{r}dz=\frac{\tilde{m}}{2\pi\tilde{\rho}\tilde{r}_{c}\tilde{l}^{*}\tilde{d}_{r}^{*}},\label{Conservation law 0}
\end{equation}
which using (\ref{Normalized d_r and d_z}) can be rewritten as 
\begin{equation}
\int_{0}^{\infty}pdz=\frac{\tilde{m}}{\pi\left(2-\nu\right)\tilde{\rho}\tilde{r}_{c}\tilde{l}^{*}\tilde{d}_{r}^{*}}.\label{Conservation law}
\end{equation}
In the following, we assume that the fluid mass, $\tilde{m}$, is
such that the right-hand side of (\ref{Conservation law}) is unity,
$\int_{0}^{\infty}pdz=1$. Note that the conservation law, (\ref{Conservation law}),
can be also derived by integrating (\ref{Evolution eqaution for the pressure})
with respect to $z$ from $0$ to $\infty$, using (\ref{IC impulse}).
Following Di Federico and Ciriello \cite{DiFederico}, who focused on shear-thinning fluids and considered
an instantaneous mass injection of a weakly compressible power-law fluid
into porous medium, we construct self-similar solutions by defining
a similarity variable $\xi$ 
\begin{equation}
\xi=zt^{-\frac{n}{2}},\label{Similarity variable}
\end{equation}
and self-similar profile $f$
\begin{equation}
p(z,t)=t^{-\frac{n}{2}}f(\xi).\label{Pressure self smilar form}
\end{equation}
For a Newtonian fluid ($n=1$) the solution is well-known 
\begin{equation}
p(z,t)=\frac{1}{\sqrt{\pi t}}\exp\left(-\frac{z^{2}}{4t}\right)=t^{-1/2}\frac{1}{\sqrt{\pi}}\exp\left(-\frac{\xi^{2}}{4}\right)\triangleq t^{-\frac{1}{2}}f(\xi).\label{Pressure self smilar form-1}
\end{equation}
For non-Newtonian fluids, substituting (\ref{Pressure self smilar form}) into (\ref{Evolution eqaution for the pressure})
and (\ref{Conservation law}) yields an ordinary differential equation
for the self-similar profile and the corresponding conservation law,
\begin{equation}
\frac{d}{d\xi}\left(\left|\frac{df(\xi)}{d\xi}\right|^{\frac{1}{n}-1}\frac{df(\xi)}{d\xi}+\frac{n}{2}\xi f(\xi)\right)=0,\label{ODE}
\end{equation}
and
\begin{equation}
\int_{0}^{\infty}f(\xi)d\xi=1.\label{Conservation law 1}
\end{equation}
Integrating (\ref{ODE}) with respect to $\xi$ and applying the far-field
condition of vanishing pressure and its derivatives $f=df/d\xi\rightarrow0$
as $\xi\rightarrow\infty$, yield
\begin{equation}
\left|\frac{df(\xi)}{d\xi}\right|^{\frac{1}{n}-1}\frac{df(\xi)}{d\xi}+\frac{n}{2}\xi f(\xi)=0.\label{ODE-2}
\end{equation}
Since, the pressure should be positive and the pressure gradient
should be negative, i.e. $f\geq0$ and $df/d\xi\leq0$, we can drop the absolute value in (\ref{ODE-2})
and obtain 
\begin{equation}
\frac{df}{d\xi}=-\left(\frac{n}{2}\right)^{n}\xi^{n}f(\xi)^{n}<0.\label{ODE-3}
\end{equation}
Integrating (\ref{ODE-3}) with respect to $\xi$ leads to
\begin{equation}
f(\xi)=\xi_{*}^{\frac{1+n}{1-n}}\left|k\right|^{\frac{1}{1-n}}\left(1\mp\frac{\xi^{n+1}}{\xi_{*}^{n+1}}\right)_{+}^{\frac{1}{1-n}},\label{Self-similar profile Impulse}
\end{equation}
where the minus and plus signs stand for $n<1$ and $n>1$, respectively,
$k$ is defined as
\begin{equation}
k=\left(\frac{n}{2}\right)^{n}\frac{1-n}{1+n},\label{k Impulse}
\end{equation}
$x_{+}=\max\left\{ x,0\right\} $ and $\xi_{*}$ is a positive constant
uniquely determined from (\ref{Conservation law 1}) and given by
\begin{equation}
\xi_{*}=\left\{\begin{array}{ll}
\xi_{front}=\left(\frac{1}{\left|k\right|^{\frac{1}{1-n}}}\frac{\Gamma(\frac{3-n^{2}}{1-n^{2}})}{\Gamma(\frac{2-n}{1-n})\Gamma(\frac{2+n}{1+n})}\right)^{\frac{1-n}{2}} & n<1\\
\xi_{no\,front}=\left(\frac{1}{\left|k\right|^{\frac{1}{1-n}}}\frac{\Gamma(\frac{1}{n-1})}{\Gamma(\frac{2}{n^{2}-1})\Gamma(\frac{2+n}{1+n})}\right)^{\frac{1-n}{2}} & n>1
\end{array},\label{front Impulse} \right.
\end{equation}
where $\Gamma$ is the Gamma function \cite{Abramowitz}.

For a shear-thinning fluid, $n<1$, the self-similar profile $f(\xi)$ becomes uniformly zero for  $\xi>\xi_{front}$ thus indicating the existence of a front which moves according to $z_{front}(t)=\xi_{front}t^{\frac{n}{2}}$. The condition $f=df/d\xi\rightarrow0$
holds now at $\xi=\xi_{front}$. In contrast to the shear-thinning fluid,
the self-similar profile $f(\xi)$ of the shear-thickening fluid, $n>1$,
approaches zero only as $\xi\rightarrow\infty$, similarly to the case of a Newtonian fluid.

The closed-form solutions,
(\ref{Self-similar profile Impulse}), provide the associated radial deformation
\begin{equation}
d_{r}(z,t)=\left(1-\frac{\nu}{2}\right)p(z,t)=\left(1-\frac{\nu}{2}\right)t^{-\frac{n}{2}}f(\xi),\label{d_r}
\end{equation}
which admits the same self-similar profile $f$. Substituting (\ref{Pressure self smilar form-1})
and (\ref{Self-similar profile Impulse}) into (\ref{Normalized d_r and d_z}),
we obtain the corresponding self-similar axial deformation in the
form $d_{z}(z,t)=\left(0.5-\nu\right)\int_{0}^{z}p(z,t)dz=\left(0.5-\nu\right)D_{z}(\xi)$,
where $D_{z}(\xi)$ is expressed in terms of generalized hyper-geometric functions
\begin{figure}
 \centerline{\includegraphics[scale=1.2]{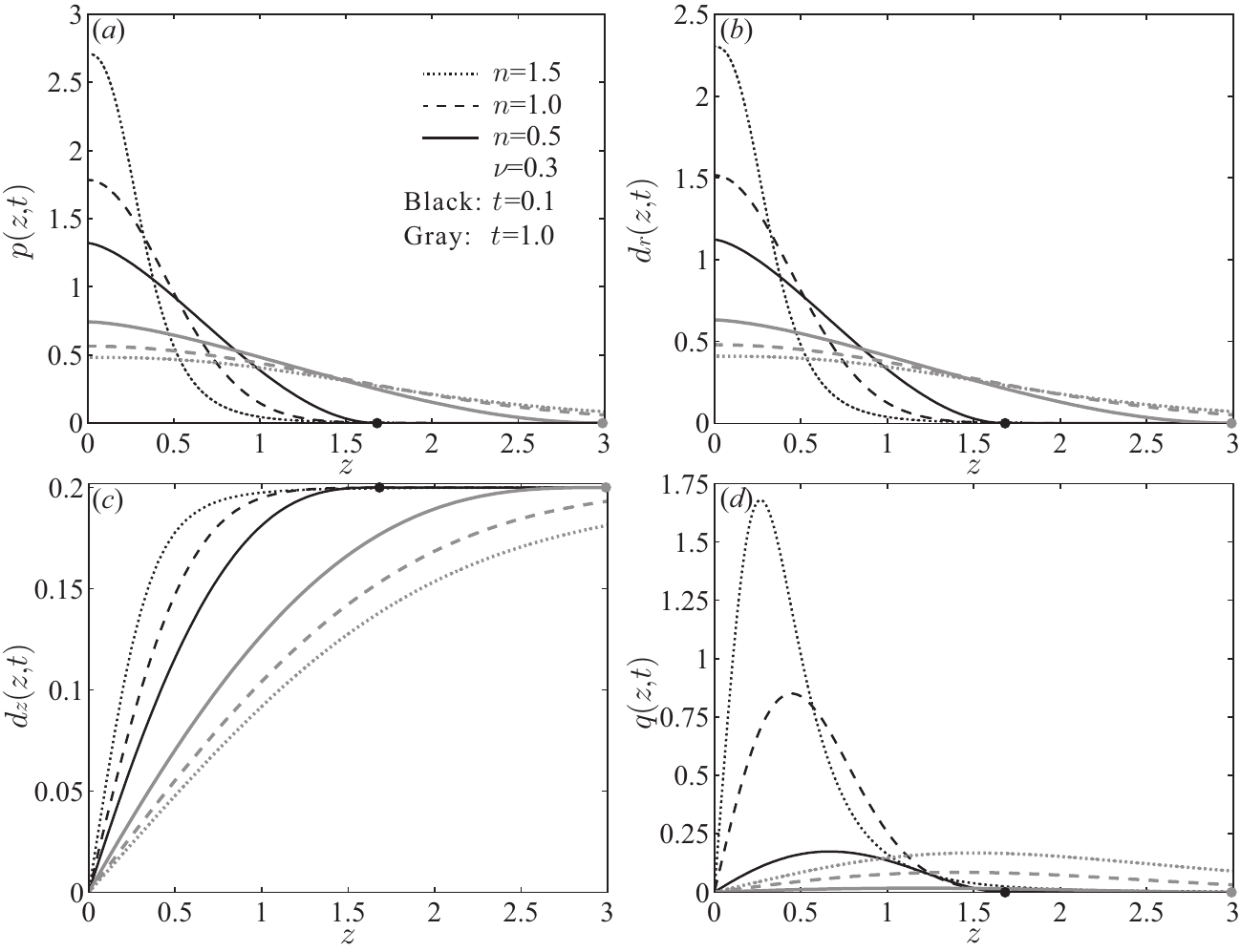}}
\caption{Propagation of a power-law fluid in an elastic cylinder due to instantaneous injection of mass at $z=0$. (a) and (b) present, respectively, the pressure distribution and radial deformation as a function of the axial coordinate, both of which are characterized by the decay rate of  $t^{-n/2}$. (c) and (d) present, respectively, the axial deformation and volume flux as a function of the axial coordinate, both of which have a decay rate that is independent of $n$. Black and gray lines correspond to $t=0.1$ and $t=1.0$,
respectively. Dotted lines represent a shear-thickening fluid ($n=1.5)$,
dashed lines represent a Newtonian fluid ($n=1)$ and solid lines represent
a shear-thinning fluid ($n=0.5)$. At early times the shear-thickening fluid is more localized (but not compactly supported) than the Newtonian and shear-thinning fluids, while for longer times the shear-thinning fluid has the lowest propagation rate remaining compactly supported. Black and gray dots indicate the location of the front of shear-thinning fluid, $z_{front}(t)=2.99t^{1/4}$.}
\end{figure}
\begin{equation}
D_{z}(\xi)=\left\{\begin{array}{ll}
\xi_{*}^{\frac{1+n}{1-n}}\left|k\right|^{\frac{1}{1-n}}\xi{}_{2}F_{1}\left(\frac{1}{n-1},\frac{1}{n+1};\frac{n+2}{n+1};\pm\frac{\xi^{n+1}}{\xi_{*}^{n+1}}\right) & n\neq1\\
\mathrm{erf}\left(\frac{\xi}{2}\right) & n=1
\end{array},\label{Self-similar profile Dz Impulse} \right.
\end{equation}
where $_{2}F_{1}\left(a,b;c;z\right)$ is the Gaussian hyper-geometric
function \cite{Abramowitz,Slater}. The plus and minus signs correspond to $n<1$ and $n>1$,
respectively. Rescaling the time in the expression for the volume flux, (\ref{flux}), according to (\ref{Time rescaling}),
we find 
\begin{equation}
q=\frac{n}{3n+1}\frac{\pi}{2^{\frac{1}{n}}}\left(-\left|\frac{\partial p}{\partial z}\right|^{\frac{1}{n}-1}\frac{\partial p}{\partial z}+\frac{2}{5-4\nu}\frac{\partial d_{z}}{\partial t}\right).\label{Flux rescaled}
\end{equation}
Expressing (\ref{Flux rescaled}) in terms of the similarity variable
$\xi$ yields
\begin{equation}
q=\frac{n^{2}}{3n+1}\frac{\pi}{2^{\frac{1}{n}}}\frac{2-\nu}{5-4\nu}t^{-1}\xi f(\xi)=\frac{n^{2}}{3n+1}\frac{\pi}{2^{\frac{1}{n}}}\frac{2-\nu}{5-4\nu}t^{-1}Q(\xi),\label{Flux Impulse}
\end{equation}
where $f(\xi)$ is given in (\ref{Pressure self smilar form-1}) and
in (\ref{Self-similar profile Impulse}) for the Newtonian fluid and
the non-Newtonian fluid, respectively, and $Q(\xi)=\xi f(\xi)$.

Figure (2) summarizes the results of viscous-elastic interaction due to a pressure impulse applied at the inlet for $n=0.5,1,1.5$. All solutions corresponding to shear-thinning
fluids are compactly supported, in contrast to solutions of Newtonian
and shear-thickening fluids. Figures 2(a) and 2(b) present the pressure distribution and the radial
deformation versus the axial coordinate $z$, for different times. From (\ref{Pressure self smilar form})
and (\ref{d_r}), the decay rate in this case scales as $t^{-n/2}$,
thus indicating that for long times the pressure and the radial deformation
of shear-thickening fluids decay (and propagate) faster than in the
case of Newtonian and shear-thinning fluids, consistent with the results
of Figs. 2(a) and 2(b). Figures 2(c) and 2(d) show the axial deformation
and volume flux versus axial coordinate. In this case, the decay rate
is similar for Newtonian and non-Newtonian fluids (i.e. is independent of $n$). 

It worth to note that at early-times, $t\ll1$, the shear-thickening fluids have the lowest propagation rate and are thus more localized (but not compactly supported) than the Newtonian and shear-thinning fluids. This is inverted at longer times as propagation rate of the shear-thickening fluids increases, and they are least localized. 

Furthermore, we note that the current analysis is limited to an intermediate range of shear rates in accordance the validity range of the power-law model. For sufficiently long times, as $\partial p/ \partial z$ diminishes, the shear rate will also diminish, driving the effective viscosity outside the validity range of the power-law model. Thus, the current analysis will become invalid at long times. In practice, in such cases the effective viscosity would approach a constant value, and the dynamics could be described by a linear diffusion equation.

\subsection{Sudden change of inlet pressure}

We here examine the viscous-elastic dynamics associated with a sudden
change of inlet pressure 
\begin{equation}
p(z=0,t)=p_{0}H(t),\label{BC p_in-1}
\end{equation}
from an initially uniform pressure within the cylinder, $p(z,t=0)=p_{i}$.
Here $H$ stands for the Heaviside function, whereas $p_{i}$ and
$p_{0}$ are the magnitudes of the initial and the applied pressure, respectively. 
The pressure gradient is positive for $p_{i}>p_{0}$ and negative
for $p_{i}<p_{0}$, corresponding to filling or discharging
of the elastic tube, respectively. 

For convenience, we define $\beta=\mathrm{sgn}\left(p_{i}-p_{0}\right)$
and rewrite (\ref{Evolution eqaution for the pressure}) as 
\begin{equation}
\frac{\partial p}{\partial t}-\frac{\partial}{\partial z}\left(\left(\beta\frac{\partial p}{\partial z}\right)^{\frac{1}{n}-1}\frac{\partial p}{\partial z}\right)=0.\label{Evolution equation for the pressure with beta}
\end{equation}
Following Pascal and Pascal \cite{Pascal1985}, we define a similarity variable $\xi$ 
\begin{equation}
\xi=zt^{-\frac{n}{n+1}},\label{Similarity variable 2}
\end{equation}
and require 
\begin{equation}
p(z,t)=p(\xi).\label{Pressure self similar p_in}
\end{equation}
The corresponding solution for a Newtonian fluid solution is 
\begin{equation}
p(z,t)=p_{0}+\left(p_{i}-p_{0}\right)\mathrm{erf}\left(\frac{z}{2\sqrt{t}}\right),\quad p(\xi)=p_{0}+\left(p_{i}-p_{0}\right)\mathrm{erf}\left(\frac{\xi}{2}\right).\label{Newtonian solution p_in}
\end{equation}
For the non-Newtonian case, we substitute (\ref{Similarity variable 2})
and (\ref{Pressure self similar p_in}) into (\ref{Evolution equation for the pressure with beta}),
obtaining an ordinary differential equation for the pressure 
\begin{equation}
\frac{d^{2}p(\xi)}{d\xi^{2}}+\beta\frac{n^{2}}{n+1}\xi\left(\beta\frac{dp(\xi)}{d\xi}\right)^{\frac{2n-1}{n}}=0,\label{ODE p_in}
\end{equation}
with boundary conditions for $t>0$
\begin{equation}
p(\xi=0)=p_{0},\quad p(\xi\rightarrow\infty)=p_{i},\quad\frac{dp}{d\xi}(\xi\rightarrow\infty)=0.\label{BC p_in}
\end{equation}
Integrating once (\ref{ODE p_in}) with respect to $\xi$ yields
\begin{equation}
\frac{dp}{d\xi}=\beta\xi_{*}^{\frac{2n}{1-n}}\left|k\right|^{\frac{n}{1-n}}\left(1\mp\frac{\xi^{2}}{\xi_{*}^{2}}\right)_{+}^{\frac{n}{1-n}},\label{dp/dxi p_in}
\end{equation}
where the upper and lower signs correspond to $n<1$ and $n>1$, respectively. The parameter
$k$ is defined as
\begin{equation}
k=\frac{n}{2}\frac{1-n}{1+n},\label{k p_in}
\end{equation}
and $\xi_{*}$ is a constant to be determined from boundary conditions on the pressure, in contrast to Sec. IV A where a similar constant is found from conservation of mass.

Integrating (\ref{dp/dxi p_in}) with respect to $\xi$ and applying the
boundary condition (\ref{BC p_in}) at $\xi=0$, leads to
\begin{equation}
p(\xi)=p_{0}+\beta\xi_{*}^{\frac{2n}{1-n}}\left|k\right|^{\frac{n}{1-n}}\xi{}_{2}F_{1}\left(\frac{1}{2},\frac{n}{n-1};\frac{3}{2};\pm\frac{\xi^{2}}{\xi_{*}^{2}}\right).\label{pressure p_in}
\end{equation}
To determine the constant $\xi_{*}$ for different values of $n$, we utilize the boundary condition $p(\xi\rightarrow\infty)=0$,
yielding
\begin{equation}
\xi_{*}=\left\{\begin{array}{ll}
\xi_{front}=\left(\frac{2\left|p_{i}-p_{0}\right|}{\sqrt{\pi}\left|k\right|^{\frac{n}{1-n}}}\frac{\Gamma(\frac{n-3}{2(n-1)})}{\Gamma(\frac{1}{1-n})}\right)^{\frac{1-n}{1+n}} & n<1\\
\xi_{no\,front}=\left(\frac{2\left|p_{i}-p_{0}\right|}{\sqrt{\pi}\left|k\right|^{\frac{n}{1-n}}}\frac{\Gamma(\frac{n}{(n-1)})}{\Gamma(\frac{n+1}{2(n-1)})}\right)^{\frac{1-n}{1+n}} & n>1
\end{array}.\label{front p_in} \right.
\end{equation}
From (\ref{dp/dxi p_in}) we obtain that for shear-thinning fluids
$dp/d\xi=0$ at $\xi=\xi_{*}=\xi_{front}$, thus indicating the existence of a front
moving according to $z_{front}(t)=\xi_{front}t^{n/(n+1)}$. On the other hand, for
shear-thickening fluids it follows that $dp/d\xi=0$ only as $\xi\rightarrow\infty$,
and therefore no front exists.

Figures 3(a) and 3(b) present the resulting pressure distribution,
(\ref{pressure p_in}), and the radial deformation,
$d_{r}(z,t)=\left(1-0.5\nu\right)p(z,t)=\left(1-0.5\nu\right)p(\xi)$, corresponding to the case of $p_{0}=1,\,p_{i}=0,\,\beta=-1$,
for different times. Figure 3(c) shows the axial deformation, $d_{z}(z,t)$,
given as $d_{z}(z,t)=\left(0.5-\nu\right)\int_{0}^{z}p(z,t)dz=\left(0.5-\nu\right)zD_{z}(\xi)/\xi$, where for the Newtonian fluid ($n=1$) $D_{z}(\xi)$ is the following self-similar profile,
\begin{equation}
D_{z}(\xi)=p_{0}\xi+\left(p_{i}-p_{0}\right)\left(\frac{2}{\sqrt{\pi}}\left(-1+\mathrm{e}^{-\xi^{2}/4}\right)+\xi\mathrm{erf}\left(\frac{\xi}{2}\right)\right),\label{D_z Newtonian p_in}
\end{equation}
and for a non-Newtonian fluid is given by
\begin{eqnarray}
D_{z}(\xi)& = &p_{0}\xi+\beta\xi_{*}^{\frac{2n}{1-n}}\left|k\right|^{\frac{n}{1-n}}\left[\xi^{2}{}_{2}F_{1}\left(\frac{1}{2},\frac{n}{n-1};\frac{3}{2};\pm\frac{\xi^{2}}{\xi_{*}^{2}}\right)
  \right.\nonumber\\~~~~~~~~~~~~~~~~~~
&& \left.+\frac{1}{2}\left(n-1\right)\xi_{*}^{2}\left(\mp\xi^{2}+\xi_{*}^{2}\right)^{\frac{n}{1-n}}\left(\xi^{2}\xi_{*}^{\frac{2}{n-1}}\mp\xi_{*}^{\frac{2n}{n-1}}\pm\left(\mp\xi^{2}+\xi_{*}^{2}\right)^{\frac{n}{n-1}}\right) \right], 
\label{D_z non-Newtonian p_in}
\end{eqnarray}where the upper signs in (\ref{D_z non-Newtonian p_in}) correspond to $n<1$ and the lower signs correspond to $n>1$, respectively.
The solution for volume flux is presented in Fig. 3(d) and is obtained
using (\ref{Flux rescaled}),
\begin{equation}
q=\frac{n}{3n+1}\frac{\pi}{2^{\frac{1}{n}}}t^{-\frac{1}{n+1}}\left[\left(\beta\frac{dp}{d\xi}\right)^{\frac{1}{n}}+\frac{n}{n+1}\frac{1-2\nu}{5-4\nu}\left(D_{z}(\xi)-\xi\frac{dD_{z}(\xi)}{d\xi}\right)\right],\label{flux p_in}
\end{equation}
where the expressions of $p(\xi)$ and $D_{z}(\xi)$ are given in
(\ref{Newtonian solution p_in}) and (\ref{D_z Newtonian p_in}) for
the Newtonian fluid, and in (\ref{pressure p_in}) and (\ref{D_z non-Newtonian p_in})
for the non-Newtonian fluid.

\begin{figure}
 \centerline{\includegraphics[scale=1.2]{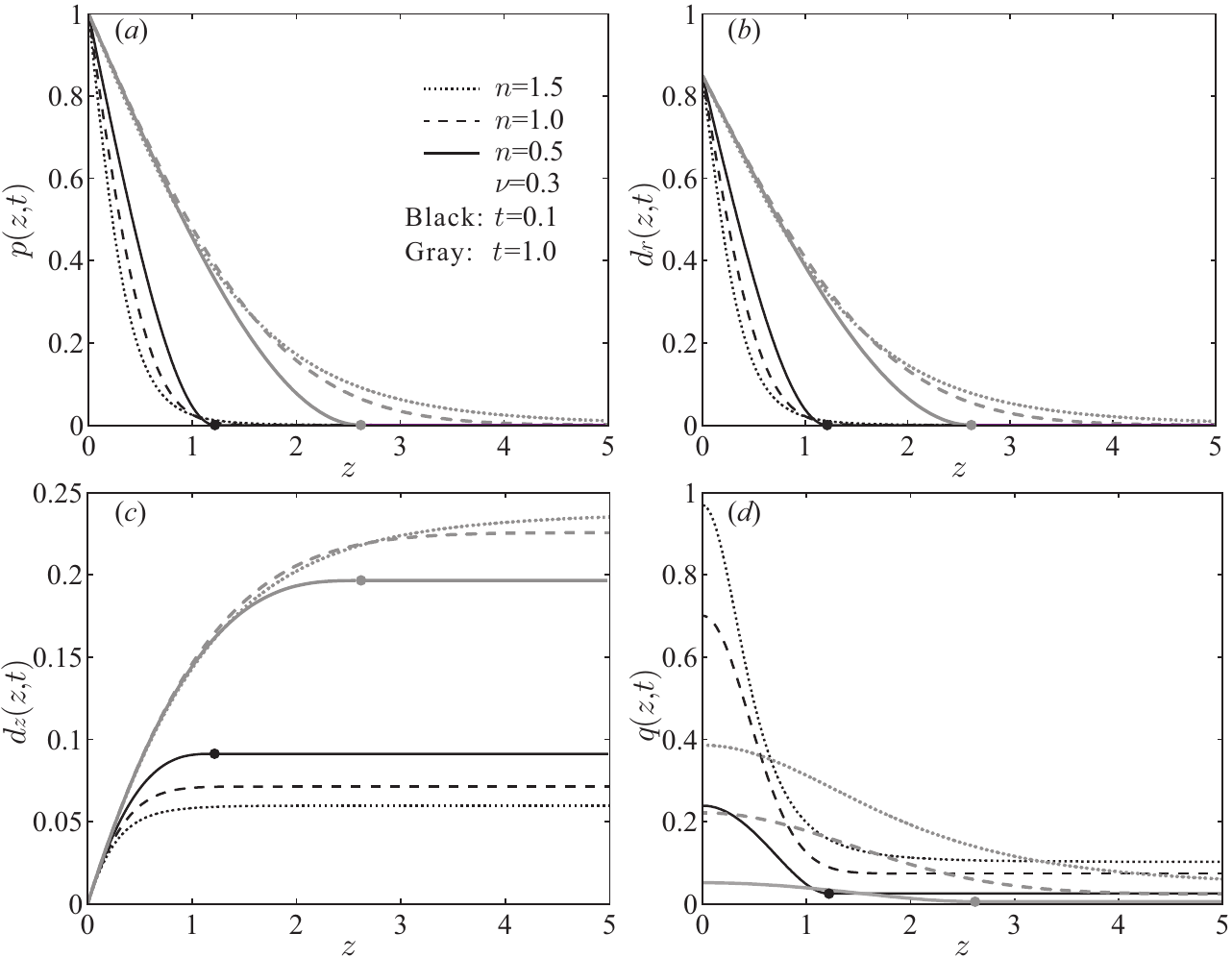}}
\caption{Viscous-elastic dynamics of power-law fluids within an elastic cylinder due to a step increase in inlet pressure. (a), (b) and (c) present, respectively, the pressure distribution and the radial and axial deformations a function of the axial coordinate. While for sufficiently long times, $t\rightarrow\infty$, the radial deformation approaches a constant value, $d_{r}\rightarrow 0.85$, the axial deformation is a linear function of $z$ with the slope of 0.2, $d_{z}/z\rightarrow 0.2$, independently of $n$. (d) shows the volume flux versus axial coordinate, characterized
by the decay rate of $t^{-1/(n+1)}$. Black and gray lines
correspond to $t=0.1$ and $t=1.0$, respectively. Dotted lines represent
a shear-thickening fluid ($n=1.5)$, dashed lines represent a Newtonian
fluid ($n=1)$ and solid lines represent a shear-thinning fluid ($n=0.5)$. Notably, self-similar solutions exhibit a compactly supported propagation front (black and gray dots),  $z_{front}(t)=2.62t^{1/3}$, solely for shear-thinning fluids. All calculations were performed
using $p_{0}=1$, $p_{i}=0$ and $\nu=0.3$.}
\end{figure}

While the Newtonian and non-Newtonian fluids have the same decay rate
in pressure and radial deformation, the shear-thinning fluids exhibit
the fastest decay in volume flux and the shear-thickening fluids exhibit
the fastest growth in axial deformation.

\subsection{Steady-state dynamics due to an oscillatory inlet pressure}

In this section, we examine the steady non-Newtonian viscous-elastic interaction resulting from an oscillatory pressure at the inlet 
\begin{equation}
p(z=0,t)=\mathrm{Re}\left\{ \mathrm{e}^{i\omega t}\right\} ,\label{BC p_Os}
\end{equation}
where $\omega$ is the non-dimensional magnitude of the base frequency
$1/\tilde{t}^{*}$, and the far-field boundary conditions is  $p(z\rightarrow\infty,t)=0$.

The corresponding steady-state solution for a Newtonian fluid is 
\begin{equation}
p(z,t)=\mathrm{e}^{-\sqrt{\omega/2}z}\mathrm{Re}\left\{ \mathrm{e}^{i\left(\omega t-\sqrt{\omega/2}z\right)}\right\} =\mathrm{e}^{-\sqrt{\omega/2}z}\cos\left(\omega t-\sqrt{\omega/2}z\right).\label{LO Solution p_Os}
\end{equation}
For a non-Newtonian fluid, an exact solution of (\ref{Evolution eqaution for the pressure}) together with the boundary condition (\ref{BC p_Os}) is difficult to obtain. In the Supplemental Material \cite{SI} we present an asymptotic solution in terms of Green's functions, but it is difficult to obtain direct insight from this solution as some of the resulting expressions can be evaluated only numerically.  
Nevertheless, some physical insight on this problem can be obtained by relying on the first-order asymptotic correction without the need for an exact solution. As opposed to Newtonian fluids where the resulting pressure oscillates only with the imposed frequency $\omega$ (see (\ref{LO Solution p_Os})), we expect non-Newtonian fluids to respond at various frequencies due to the non-linearity of  (\ref{Evolution eqaution for the pressure}). To explore these active frequencies $\omega_{ai}$, we examine the source term of the first-order equation (\ref{First order pressure}) that reflects the non-Newtonian
response of the fluid
\begin{multline}
\frac{\partial p^{(1)}}{\partial t}-\frac{\partial^{2}p^{(1)}}{\partial z^{2}} =
 -\frac{1}{2}\omega\mathrm{e}^{-\sqrt{\omega/2}z}\sin\left(\sqrt{\frac{\omega}{2}}z-\omega t\right)\\
\times\left[ 2+\ln\frac{\omega}{2}-\sqrt{2\omega}z+\ln\left(1+\sin\left(\sqrt{2\omega}z-2\omega t\right)\right) \right].
\label{FO Asym Eq-Os}
\end{multline}
By decomposing the right-hand side of (\ref{FO Asym Eq-Os}) into
three contributions:\begin{subequations}
\begin{equation}
S_{1}\left(z,t\right)=-\frac{1}{2}\omega\left(2+\ln\frac{\omega}{2}\right)\mathrm{e}^{-\sqrt{\omega/2}z}\sin\left(\sqrt{\frac{\omega}{2}}z-\omega t\right),\label{1 cont}
\end{equation}
\begin{equation}
S_{2}\left(z,t\right)=\sqrt{\frac{\omega^{3}}{2}}z\mathrm{e}^{-\sqrt{\omega/2}z}\sin\left(\sqrt{\frac{\omega}{2}}z-\omega t\right),\label{2 cont}
\end{equation}
\begin{equation}
S_{3}\left(z,t\right)=-\frac{1}{2}\omega\mathrm{e}^{-\sqrt{\omega/2}z}\ln\left(1+\sin\left(\sqrt{2\omega}z-2\omega t\right)\right)\sin\left(\sqrt{\frac{\omega}{2}}z-\omega t\right),\label{3 cont}
\end{equation}
\end{subequations}and adopting for convenience the following notations 
\begin{equation}
\sin\left[m\left(\sqrt{\frac{\omega}{2}}z-\omega t\right)\right]=\mathrm{s}\left(m\omega\right),\quad\cos\left[m\left(\sqrt{\frac{\omega}{2}}z-\omega t\right)\right]=\mathrm{c}\left(m\omega\right),\label{s}
\end{equation}
where $m=1,2,3...$, it is evident that the first two contributions (\ref{1 cont}) and
(\ref{2 cont}) act with the imposed frequency $\omega$. 

Expanding the logarithmic term in (\ref{3 cont}) in a Taylor series yields
\begin{eqnarray}
 \ln\left(1+\mathrm{s}\left(2\omega\right)\right)\mathrm{s}\left(\omega\right)&\approx&\left(1-\frac{\mathrm{s}\left(2\omega\right)}{2}+...\right)\mathrm{s}\left(\omega\right)\mathrm{s}\left(2\omega\right)=      \nonumber \\
   &=& \frac{1}{2}\left(1-\frac{\mathrm{s}\left(2\omega\right)}{2}+...\right)\left(\mathrm{c}\left(\omega\right)-\mathrm{c}\left(3\omega\right)\right)= \nonumber \\
   &=&\frac{1}{2}\left(\mathrm{c}\left(\omega\right)-\mathrm{c}\left(3\omega\right)-\frac{1}{4}\left(\mathrm{2s}\left(\omega\right)+\mathrm{s}\left(3\omega\right)-\mathrm{s}\left(5\omega\right)\right)+...\right),\label{Taylor expansion}
\end{eqnarray}
and thus we conclude that the third contribution (\ref{3 cont}) adds
additional forcing frequencies which are the multiplication of the imposed frequency with odd natural numbers ($3\omega,\,5\omega,\,...$).

To validate the Taylor-expansion approximation made in (\ref{Taylor expansion}), we performed fast Fourier transform (FFT) analysis on a numerical calculation for the source term of (\ref{FO Asym Eq-Os}). The results presented in Fig. 4(a) show full agreement with analytically predicted frequencies of the first-order source term. To further explore the viscous-elastic dynamics of the problem, we solved the governing equation (\ref{Evolution eqaution for the pressure})
numerically on a fixed domain $(0,L)$ using a second-order explicit
finite-difference scheme. The detailed description of the numerical
scheme is presented in the Supplemental Material \cite{SI}.  In all simulations we used values
of $L=10$ and $\omega=2\pi$.
\begin{figure}
 \centerline{\includegraphics[scale=1.2]{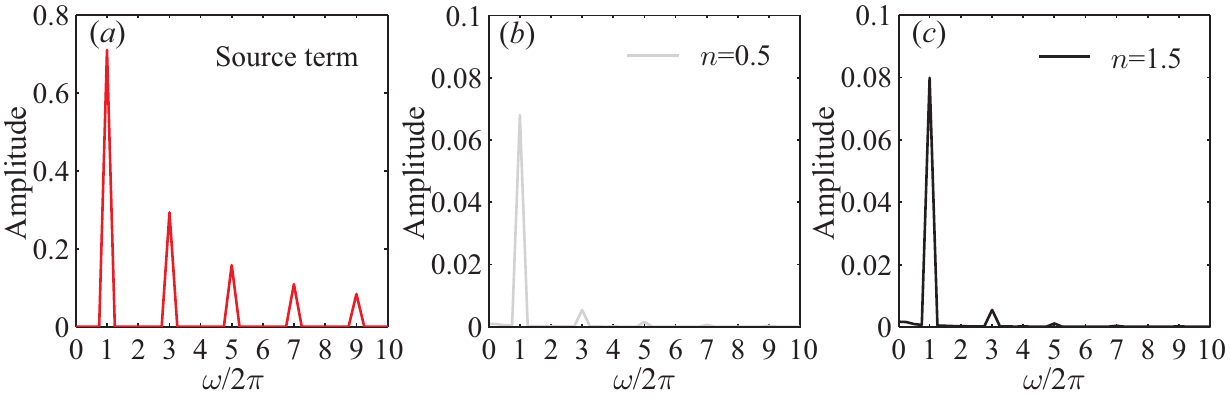}}
\caption{FFT analysis results showing active frequencies due to imposed frequency $\omega=2\pi$. (a) Resulting frequencies from FFT analysis applied directly to the source term of (\ref{FO Asym Eq-Os}). (b,c) Resulting frequencies from FFT analysis applied on the numerical solution of (\ref{Evolution eqaution for the pressure})
and corresponding to shear-thinning and shear-thickening fluids, respectively. Clearly, the source term itself predicts well the governing frequencies of the solution, though of course cannot predict the amplitudes.}
\end{figure} 

In Figs. 4(b) and 4(c) we present the resulting frequencies of FFT analysis applied on the results of numerical simulations for shear-thinning and shear-thickening fluids, showing excellent agreement with the analytically predicted frequencies obtained from the first-order source term. From Figs. 4(b) and 4(c) we obtain that for actuation at the viscous-elastic time scale ($\omega \sim O(1)$), the dominant frequencies are 
\begin{equation}
\omega_{a1}=\omega=2\pi,\quad\omega_{a2}=3\omega=3\cdot2\pi,\quad\omega_{a3}=5\omega=5\cdot2\pi,\label{active f}
\end{equation}
and therefore the steady-state  time variation of the pressure at any location
$z$ can be approximately described by the superposition of only three
sinusoidal waveforms
\begin{equation}
p(z,t)=A_{1}\left(z\right)\mathrm{e}^{i\omega_{\omega_{a1}}t}+A_{2}\left(z\right)\mathrm{e}^{i\omega_{\omega_{a2}}t}+A_{3}\left(z\right)\mathrm{e}^{i\omega_{\omega_{a3}}t}+\mathrm{c.c.},\label{Approximately three sinusoidal waveforms}
\end{equation}
where $A_{1}\left(z\right)$, $A_{2}\left(z\right)$ and $A_{3}\left(z\right)$
are complex functions solely depending on $z$ and ''c.c.'' represents
the complex conjugate of preceding terms. 

Figure (5) summarizes the steady-state results of viscous-elastic
interaction due to an oscillatory pressure (\ref{BC p_Os}). Figures
5(a,b,c) present the steady-state pressure distribution versus axial coordinate
for one period of oscillations, obtained from numerical solution of
(\ref{Evolution eqaution for the pressure}) and corresponding to
shear-thinning, Newtonian and shear-thickening fluids, respectively. Clearly, the shear-thinning fluid is characterized by the most rapid pressure decay rate similarly to the results in previous sections, thus suggesting the existence of a propagation front (which we however did not prove for this case). 
\begin{figure}
 \centerline{\includegraphics[scale=1.2]{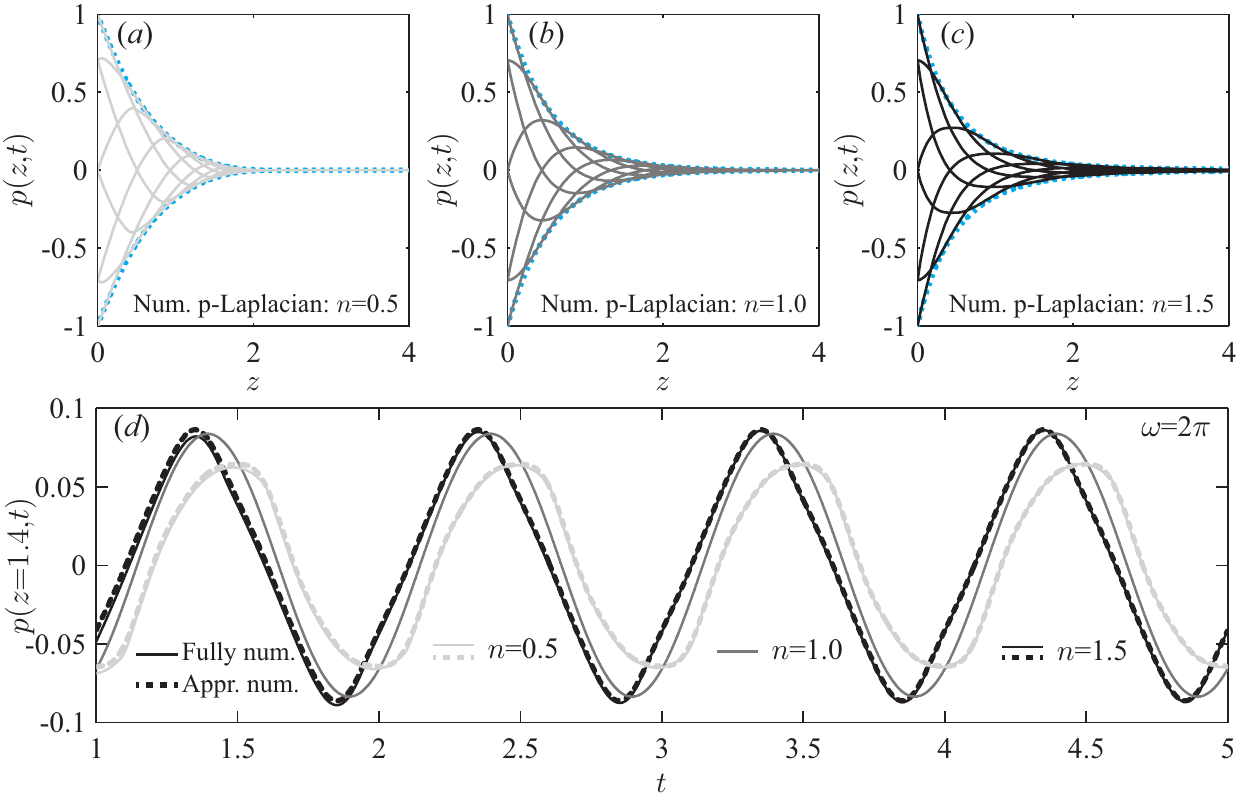}}
\caption{Numerical results showing the steady-state pressure distribution resulting
from viscous-elastic interaction induced by oscillating inlet pressure
with $\omega=2\pi$. (a,b,c) Pressure versus axial coordinate
for one time period, corresponding to shear-thinning, Newtonian, and shear-thickening
fluids, respectively. Dotted lines represent the envelope of the pressure
decay, and indicate the most rapid decay for the shear-thinning fluid. (d) Comparison of numerical
results (solid lines) and approximate solutions (dashed lines), (\ref{Approximately three sinusoidal waveforms}),
for pressure distributions at $z=1.4$ as a function of time and corresponding
to $n=0.5,1,1.5$. While Newtonian fluid (dark gray line)
is characterized by a sinusoidal waveform, shear-thinning (light gray lines) and shear-thickening (black lines) fluids exhibit non-sinusoidal ''saw-tooth''
and ''shark-tooth'' waveforms, respectively.}
\end{figure} 

In Fig. 5(d), we compare the numerical solutions (solid lines) and approximate solutions (dashed lines), (\ref{Approximately three sinusoidal waveforms}), for the pressure variation as a function of time at  $z=1.4$, showing good agreement. As can be inferred from Fig. 5(d), in the steady-state the shear-thinning fluid exhibits a non-sinusoidal ''saw-tooth'' waveform, whereas the shear-thickening is characterized by a non-sinusoidal ''shark-tooth'' waveform. 

\section{The inverse role of viscosity between viscous-elastic and Stokes' problems }

In Secs. IV B and IV C, we studied analytically and numerically the viscous-elastic
interaction of power-law fluids through an elastic cylinder resulting
from a constant or an oscillating pressure at the inlet. Specifically, we solved
a homogeneous p-Laplacian equation that can be written in dimensional
form as
\begin{equation}
\frac{\partial\tilde{p}}{\partial\tilde{t}}=\frac{n}{3n+1}\frac{\tilde{E}\tilde{h}_{m}\tilde{r}_{c}}{5-4\nu}\frac{\partial}{\partial\tilde{z}}\left(\frac{1}{\tilde{\eta}(\dot{\gamma})}\frac{\partial\tilde{p}}{\partial\tilde{z}}\right);\quad\tilde{\eta}(\dot{\gamma})=2^{\frac{1-n}{n}}\left(\frac{\tilde{\mu}_{eff}}{\tilde{r}_{c}}\right)^{\frac{1}{n}}\tilde{r}_{c}\left|\frac{\partial\tilde{p}}{\partial\tilde{z}}\right|^{\frac{n-1}{n}}.\label{p-L Our}
\end{equation}
We note that for $n=1$ (\ref{p-L Our}) reduces to an unforced version of equation (2.53) derived by Elbaz and Gat \cite{Elbaz2014}. 

This p-Laplacian equation is found in various fluid mechanics problems including Stokes' first and second problems
for power-law fluids, which involve the solution of the following
p-Laplacian equation for $\tilde{u}_{z}$
\begin{equation}
\tilde{\rho}\frac{\partial\tilde{u}_{z}}{\partial\tilde{t}}=\frac{\partial}{\partial\tilde{z}}\left(\tilde{\eta}(\dot{\gamma})\frac{\partial\tilde{u}_{z}}{\partial\tilde{z}}\right);\quad\tilde{\eta}(\dot{\gamma})=\tilde{\mu}_{eff}\left|\frac{\partial\tilde{u}_{z}}{\partial\tilde{z}}\right|^{n-1},\label{p-L Stokes}
\end{equation}
subjected to a constant or an oscillatory velocity at $\tilde{z}=0$.
The Stokes' first problem was considered by Pascal \cite{Pascal1992} who showed
that the velocity profile is compactly supported for shear-thickening
fluids. Pritchard \textit{et al.} \cite{Pritchard} presented semi-analytical, self-similar and periodic
solutions for the flow of a power-law fluid driven by a non-sinusoidal
oscillating waveform at the wall and showed that in this case these
solutions predict a finite penetration length $\tilde{l}_{p}$ for
shear-thickening fluids, beyond which the fluid is essentially unaffected
by external forcing and is motionless. To confirm these solutions, Pritchard \textit{et al.} \cite{Pritchard} studied numerically the Stokes' second problem (on
a finite numerical domain) with the sinusoidal boundary condition
and the numerical results corresponding to shear-thickening fluids
decay rapidly with the distance from the wall thus suggesting the
existence of a finite penetration length. In addition, Pritchard \textit{et al.} \cite{Pritchard}
obtained that the velocity of shear-thickening fluids adopt a ''saw-tooth''
oscillating form, while shear-thinning fluids have a ''shark-tooth''
oscillating form. 

Clearly, the results we obtained in Secs. IV B and IV C are opposite to the results of Pascal \cite{Pascal1992} and  Pritchard \textit{et al.} \cite{Pritchard}. To explain this  difference we should examine the effect of viscosity
in the p-Laplacian equations, (\ref{p-L Our}) and (\ref{p-L Stokes}),
on the fluid penetration. While in Stokes' equation, (\ref{p-L Stokes}), viscosity contributes to diffusivity and thus to fluid penetration, in the viscous-elastic equation, (\ref{p-L Our}), viscosity reduces the diffusivity and fluid penetration. This can be observed more clearly from the expressions of the penetration length, $\tilde{l}_{p}$, which can be obtained
from scaling arguments as 
\begin{equation}
\tilde{l}_{p,VE}\sim\sqrt{\tilde{E}\tilde{h}_{m}\tilde{r}_{c}\tilde{t}^{*}/\tilde{\eta}(\dot{\gamma})}\sim\sqrt{\tilde{E}\tilde{h}_{m}\tilde{t}^{*}}\frac{\left(\tilde{r}_{c}/\tilde{\mu}_{eff}\right)^{\frac{1}{2n}}}{\left|\partial\tilde{p}/\partial\tilde{z}\right|^{\frac{n-1}{2n}}},\label{Penetration length VE}
\end{equation}
\begin{equation}
\tilde{l}_{p,S}\sim\sqrt{\tilde{t}^{*}\tilde{\eta}(\dot{\gamma})/\tilde{\rho}}\sim\sqrt{\tilde{t}^{*}\tilde{\mu}_{eff}/\tilde{\rho}}\left|\frac{\partial\tilde{u}_{z}}{\partial\tilde{z}}\right|^{\frac{n-1}{2}},\label{Penetration length S}
\end{equation}
where subscript VE indicates a viscous-elastic problem parameter, S indicates a Stokes' problem parameter.
Noting that for shear-thickening fluids ($n>1$), both viscosities
(S) and (VE) tend to zero as the gradient of the velocity (or pressure)
goes to zero, it follows from (\ref{Penetration length VE}) and (\ref{Penetration length S})
that while for the Stokes' problem the penetration length, $\tilde{l}_{p,S}$,
goes to zero, for
the viscous-elastic problem the penetration length tends to infinity,
$\tilde{l}_{p,VE}\rightarrow\infty$, as $\tilde{\eta}(\dot{\gamma})\rightarrow0$. 

On the other hand, for shear-thinning fluids, both viscosities
(S) and (VE) diverge as the gradient of the velocity (or pressure)
goes to zero, and thus Stokes' problem yields an infinite penetration
length, $\tilde{l}_{p,S}\rightarrow\infty$, whereas the viscous-elastic
problem exhibits a finite penetration length indicating the existence
of the front in this case.

These conclusions are fully consistent with our results from Secs. IV B and IV C and with the results of Pascal \cite{Pascal1992} and  Pritchard \textit{et al.} \cite{Pritchard}. Shear-thinning and shear-thickening fluids play opposite roles in the Stokes' and the viscous-elastic problems.

\vspace{-0.2in}

\section{Discussion and concluding remarks}
\vspace{-0.1in}

In this work, we studied the viscous-elastic interaction and the flow of power-law fluids through an elastic cylinder. Applying the lubrication and the elastic shell approximations, we derived a p-Laplacian governing equation relating the fluidic pressure and the external forces acting on the cylinder. We showed that in this equation viscosity acts as an inverse diffusion coefficient, such that the familiar roles of shear-thinning and shear-thickening are inversed relative to similar equations that appear in Stokes' problems. For example, for instantaneous mass injection and a step pressure applied at the inlet, our analysis revealed that a compactly supported propagation front exists solely for shear-thinning fluids, in contrast to Stokes' problem where a compactly supported front is obtained for shear-thickening fluids. Furthermore, the scaling for penetration length, (\ref{Penetration length VE}), based on a power-law viscosity model, further suggests that any Dirichlet-type boundary condition imposed at $z=0$, would result in compactly supported propagation for shear-thinning fluids, consistent with our numerical observations in Sec. IV C. 

Throughout this work we have mainly considered the viscous-elastic dynamics due to
imposed pressure at the inlet. However, we note that the solutions
obtained in Sec. IV can be used to construct new solutions
resulting from an external pressure distribution $p_{e}(z,t)$. 
Applying the Laplace transform on (\ref{Evolution eqaution for the pressure})
we obtain that the solution for the pressure in Sec. IV A for impulse pressure applied at the inlet of a semi-infinite cylinder also holds for the case of a sudden localized pressure, $p_{e}=p_{e}H(t)\delta(z)$, applied to an infinitely long cylinder. Similarly, the solution for the pressure in the case of a discharging elastic tube ((\ref{pressure p_in}) with $\beta>0$,
$p_{0}=0$) also holds for discharge due to suddenly applied constant pressure $p_{e}=p_{e}H(t)$. While the solutions for the pressure in these case are identical, the solutions for the corresponding deformation fields and the volume fluxes are different, and can be obtained from (\ref{Normalized d_r and d_z})
and (\ref{Flux rescaled}). 

For weakly non-Newtonian fluids, we obtained the leading- and first-order
asymptotic approximations, and showed (see Sec. IV C) that
the source term in first-order equation, (\ref{First order pressure}), is important as it enables physical insight on the non-Newtonian behavior without a direct solution. 

The regime of an incompressible fluid with negligible inertia is commonly encountered in lab-on-chip and soft robotics applications.  For the former, controlling the propagation of mass flux within microfluidic channels could be a mechanism the timing of simple delivery on chip, or controlling the extent of mixing between reagents. For soft robotics, control with non-Newtonian liquids holds to potential for additional degrees of freedom in modulation, and particularly with the use of compactly supported deformations where part of the structure can be guaranteed to stay stationary while another is translating. Furthermore, we believe that the dependence of the front on time and on specific value of $n$, can be used to indirectly measure the rheological properties of power-law shear-thinning liquids through measurements of deformation and front location. 

%\vspace{-0.25in}

\section*{Acknowledgments}
%\vspace{-0.2in}
This project has received funding from the European Research Council (ERC) under the European Union's Horizon 2020 Research and Innovation Programme, grant agreement no. 678734 (MetamorphChip). We gratefully acknowledge support by the Israel Science Foundation (grant no. 818/13). E.B. is supported by the Adams Fellowship Program of the Israel Academy of Sciences and Humanities.

\makeatletter
\newpage

%\clearpage 


\begin{thebibliography}{0}
\expandafter\ifx\csname natexlab\endcsname\relax\def\natexlab#1{#1}\fi
\expandafter\ifx\csname bibnamefont\endcsname\relax
  \def\bibnamefont#1{#1}\fi
\expandafter\ifx\csname bibfnamefont\endcsname\relax
  \def\bibfnamefont#1{#1}\fi
\expandafter\ifx\csname citenamefont\endcsname\relax
  \def\citenamefont#1{#1}\fi
\expandafter\ifx\csname url\endcsname\relax
  \def\url#1{\texttt{#1}}\fi
\expandafter\ifx\csname urlprefix\endcsname\relax\def\urlprefix{URL }\fi
\providecommand{\bibinfo}[2]{#2}
\providecommand{\eprint}[2][]{\url{#2}}

\end{thebibliography}


\begin{thebibliography}{9}

\bibitem{DupratStone} C. Duprat and H. A. Stone (eds.), \textit{Fluid-Structure Interactions in Low-Reynolds-Number Flows}, RSC Soft
Matter Series (Royal Society of Chemistry, Cambridge, UK, 2016).

\bibitem{Ku} D. N. Ku, Blood flow in arteries, Annu. Rev. Fluid Mech. \textbf{29}, 399 (1997).

\bibitem{Nichols} W. Nichols, M. O'Rourke, and C. Vlachopoulos, \textit{McDonald's blood flow in arteries:theoretical, experimental and clinical principles} (CRC Press, 2011).

\bibitem{Hewitt2015} I. J. Hewitt, N. J. Balmforth, and J. R. De Bruyn, Elastic-plated gravity currents, Eur. J. Appl. Math. \textbf{26}, 1 (2015).

\bibitem{Pihler-Puzovic2012} D. Pihler-Puzovic, P. Illien, M. Heil, and A. Juel, Suppression of complex fingerlike
patterns at the interface between air and a viscous fluid by elastic membranes, Phys. Rev. Lett. \textbf{108}, 074502 (2012).

\bibitem{Pihler-Puzovic2013} D. Pihler-Puzovic, R. Perillat, M. Russell, A. Juel, and M. Heil, Modelling the
suppression of viscous fingering in elastic-walled Hele-Shaw cells, J. Fluid Mech. \textbf{731}, 162 (2013).

\bibitem{AlHousseiny} T. T. Al-Housseiny, I. C. Christov, and H. A. Stone, Two-phase fluid displacement and interfacial instabilities under elastic membranes, Phys. Rev. Lett. \textbf{111}, 034502 (2013).

\bibitem{Pihler-Puzovic2014} D. Pihler-Puzovic, A. Juel, and M. Heil, The interaction between viscous fingering and
wrinkling in elastic-walled Hele-Shaw cells, Phys. Fluids. \textbf{26}, 022102 (2014).

\bibitem{Pihler-Puzovic2015} D. Pihler-Puzovic, A. Juel, G. G. Peng, J. R. Lister, and M. Heil, Displacement flows under elastic membranes. Part 1. Experiments and direct numerical simulations, J. Fluid Mech. \textbf{784}, 487 (2015).

\bibitem{Peng2015} G. G. Peng, D. Pihler-Puzovic, A. Juel, M. Heil, and J. R. Lister, Displacement flows under elastic membranes. Part 2. Analysis of interfacial effects., J. Fluid Mech. \textbf{784}, 512 (2015).

\bibitem{Hosoi2004} A. E. Hosoi and L. Mahadevan, Peeling, healing, and bursting in a lubricated elastic sheet, Phys. Rev. Lett. \textbf{93}, 137802 (2004).

\bibitem{Lister2013} J. R. Lister, G. G. Peng, and J. A. Neufeld, Viscous Control of Peeling an Elastic Sheet by Bending and Pulling, Phys. Rev. Lett. \textbf{111}, 154501 (2013).

\bibitem{Holmes} D. P. Holmes, B. Tavakol, G. Froehlicher, and H. A. Stone, Control and manipulation of microfluidic flow via elastic deformations, Soft Matter \textbf{9}, 7049 (2013).

\bibitem{Shepherd} R. F. Shepherd, F. Ilievski, W. Choi, S. A. Morin, A. A. Stokes, A. D. Mazzeo, X. Chen, M. Wang, and G. M. Whitesides, Multigait soft robot, Proc. Natl Acad. Sci. \textbf{108}, 20400 (2011).

\bibitem{Martinez} R. V. Martinez, C. R. Fish, X. Chen, and G. M. Whitesides, Elastomeric origami: Programmable paper-elastomer composites as pneumatic actuators, Adv. Mater. \textbf{22}, 1376 (2012).

\bibitem{Marchese} A. D. Marchese, C. D. Onal, and D. Rus, Autonomous soft robotic fish capable of escape maneuvers using fluidic elastomer actuators, Soft Robotics \textbf{1}, 75 (2014).

\bibitem{Matia} Y. Matia, T. Elimelech, and A. D. Gat, Leveraging nternal viscous flow to extend the capabilities of beam-shaped soft robotic actuators, Soft Robotics \textbf{4}, 126 (2017).

\bibitem{Canic} S. Canic and A. Mikelic, Effective equations modeling the flow of a viscous incompressible fluid through a long elastic tube arising in the study of blood flow through small arteries, SIAM J. Appl. Dyn. Syst. \textbf{2}, 431 (2003).

\bibitem{Grotberg} J. B. Grotberg and O. E. Jensen, Biofluid mechanics in flexible tubes, Annu. Rev. Fluid Mech. \textbf{36}, 121 (2004).

\bibitem{Heil} M. Heil and A. L. Hazel, Fluid-structure interaction in internal physiological flows, Annu. Rev. Fluid Mech. \textbf{43}, 141 (2011).

\bibitem{Elbaz2014} S. B. Elbaz and A. D. Gat, Dynamics of viscous liquid within a closed elastic cylinder subject to external forces with application to soft robotics, J. Fluid Mech. \textbf{758}, 221 (2014).

\bibitem{Hardy} B. S. Hardy, K. Uechi, J. Zhen, and H. P. Kavehpour, The deformation of flexible PDMS microchannels under a pressure driven flow, Lab on a Chip \textbf{9}, 935 (2009).

\bibitem{deRutte} J. M. de Rutte, K. G. H. Janssen, N. R. Tas, J. C. T. Eijkel, and S. Pennathur, Numerical investigation of micro-and nanochannel deformation due to discontinuous electroosmotic flow, Microfluid Nanofluid \textbf{20}, 150 (2016).

\bibitem{Pascal1992} H. Pascal, Similarity solutions to some unsteady flows of non-Newtonian fluids of power-law behavior,  Intl J. Non-Linear Mech. \textbf{27}, 759 (1992).

\bibitem{Ai} L. Ai and K. Vafai, An investigation of Stokes' second problem for non-Newtonian fluids, Numer. Heat Tr. A-Appl. \textbf{47}, 955 (2005).

\bibitem{Pritchard} D. Pritchard, C. R. McArdle, and S. K. Wilson, The Stokes boundary layer for a power-law fluid, J. Non-Newtonian Fluid Mech. \textbf{166}, 745 (2011).

\bibitem{Bird} R. B. Bird, R. C. Armstrong, and O. Hassager, \textit{Dynamics of polymeric liquids. Volume 1: Fluid Mechanics} (2nd ed. John Wiley and Sons, 1987).

\bibitem{Timoshenko} S. Timoshenko and J. N. Goodier, \textit{Theory of elasticity} (McGraw-Hill, New York, 1951).

\bibitem{Howell} P. Howell, G. Kozyreff, and J. Ockendon, \textit{Applied solid mechanics} (Cambridge University Press, 2009).

\bibitem{SI} See Supplemental Material for additional details.

\bibitem{Pascal1985} H. Pascal, and F. Pascal, Flow of non-Newtonian fluid through porous media, Intl J. Engng Sci. \textbf{23}, 571 (1985).

\bibitem{Pascal1991} H. Pascal, On propagation of pressure disturbances in a non-Newtonian fluid flowing through a porous medium,  Intl J. Non-Linear Mech. \textbf{26}, 475 (1991).

\bibitem{DiFederico} V. Di Federico and V. Ciriello, Generalized solution for 1-d non-Newtonian flow in a porous domain due to an instantaneous mass injection,  Transp. Porous Med. \textbf{93}, 63 (2012).

\bibitem{Duffy} B. R. Duffy, D. Pritchard and S. K. Wilson, The shear-driven Rayleigh problem for generalised Newtonian fluids, J. Non-Newtonian Fluid Mech. \textbf{206}, 11 (2014).

\bibitem{Ross} A. B. Ross, S. K. Wilson, and B. R. Duffy, Blade coating of a power-law fluid, Phys. Fluids. \textbf{11}, 958 (1999).

\bibitem{Cheng} Q. Cheng, Z. Sun, G. A. Meininger, and M. Almasri,  Note: Mechanical study of micromachined polydimethylsiloxane elastic microposts, Rev. Sci. Instrum. \textbf{81}, 106104 (2010).

\bibitem{Longo} S. Longo, V. Di Federico, and V. Chiapponi, A dipole solution for power-law gravity currents in porous formations, J. Fluid Mech. \textbf{778}, 534 (2015).

\bibitem{Abramowitz} M. Abramowitz and I. A. Stegun, \textit{Handbook of Mathematical Functions} (Dover, 1964).

\bibitem{Slater} L. J. Slater, \textit{Generalized hypergeometric functions} (Cambridge University Press, 1966).

\end{thebibliography}
\end{document}